# Ownership and Agency of an Independent Supernumerary Hand Induced by an Imitation Brain-Computer Interface

Luke Bashford[1&2*] and Carsten Mehring[2]

[1]Department of Bioengineering, Imperial College London, London, UK

[2]Bernstein Centre Freiburg, University of Freiburg, Freiburg, Germany

* Corresponding author

Email: luke.bashford11@imperial.ac.uk (LB)

# Abstract

To study body ownership and control, illusions that elicit these feelings in non-body objects are widely used. Classically introduced with the Rubber Hand Illusion, these illusions have been replicated more recently in virtual reality and by using brain-computer interfaces. Traditionally these illusions investigate the replacement of a body part by an artificial counterpart, however as brain-computer interface research develops it offers us the possibility to explore the case where non-body objects are controlled in addition to movements of our own limbs. Therefore we propose a new illusion designed to test the feeling of ownership and control of an independent supernumerary hand. Subjects are under the impression they control a virtual reality hand via a brain-computer interface, but in reality there is no causal connection between brain activity and virtual hand movement but correct movements are observed with 80% probability. These imitation brain-computer interface trials are interspersed with movements in both the subjects' real hands, which are in view throughout the experiment. We show that subjects develop strong feelings of ownership and control over the third hand, despite only receiving visual feedback with no causal link to the actual brain signals. Our illusion is crucially different from previously reported studies as we demonstrate independent ownership and control of the third hand without loss of ownership in the real hands.

# Introduction

Illusions that create the sense of ownership for non-body objects and agency for their actions are used to investigate the neural mechanisms underpinning the above mentioned senses of the self. The classic rubber hand illusion is one common example [1]; in this illusion the experimenter brushes a rubber hand that has been placed on a table in the position of the subjects real right hand. Synchronously the experimenter brushes the subject's real hand, which is hidden from view. This leads subjects to report that they perceive the rubber hand as though it were their real hand. Illusions of ownership have also been created from more recent studies with brain-computer interface users [2] and subjects seeing virtual reality [3,4] experiencing the feeling of ownership over the artificial devices and agency of their potential or actual movements.

In many studies the feeling of ownership for the non-body object occurs as it replaces the body part which is hidden from view. It has been demonstrated however that the brain is also capable of extending the body image beyond the physical limits of the human form. Supernumerary limb illusions demonstrate the brains ability to feel ownership over an additional limb [5–7] however in such cases it appears as though the feeling of ownership is more that of a replication of the existing limb rather than the addition of a truly independent supernumerary limb, but this distinction is not made specifically in these studies.

The feeling of an independent supernumerary limb has been presented in a medical case study [8], but recent advances in robotics and computer science suggest that artificial supernumerary limbs could be incorporated into healthy human function in the future. Supernumerary robotic limbs can be built to assist human movements [9,10] and brain-computer interfaces have been shown to work during concurrent movement execution [11]. It is therefore important to understand if the brain is capable of feeling ownership over additional and independent supernumerary limbs and agency over their actions. We therefore present here a new body ownership illusion generated by an imitation brain-computer interface, producing such feelings of ownership and agency. In our study subjects imagine moving a third virtual reality (VR) arm, while they are under the belief they control the movements of the VR arm using a brain-computer interface. We compare the strength of this illusion using both questionnaires and physiological measures: galvanic skin response (GSR) to measure ownership, hand skin temperature to measure disownership and questionnaire data to measure the feeling of ownership and agency. Furthermore we compare our illusion to the rubber hand illusions and the third rubber hand illusion. In particular we highlight the differences in terms of independence of the ownership and agency between our and the previous illusions.

# Materials and Methods

36 subjects participated in this study (16 female, 20 male, age 19-40, mean 26). All subjects performed the imitation of a BCI third hand (IBCI) and the rubber hand illusion (RHI) experiments, a subset of 17 participants additionally performed the rubber third hand illusion (RTHI) experiment and answered questions 13 and 14 for all experimental conditions. A further subset of 8 participants additionally performed a control experiment for each illusion condition. These control conditions were performed to establish a baseline measure of responses to assess the presence and strength of the illusions in the 'illusion conditions'. In the following these control conditions will, therefore, be called baseline conditions or baseline experiments. All subjects signed informed consent for participation and the experimental procedures were approved by the ethics committee at the University of Freiburg, Germany.

## Imitation of BCI Third Hand - IBCI

Subjects were sat with only their hands visible on a table top and between the subjects own hands a virtual reality hand was projected, a realistically sized left hand shown facing palm up (Fig 1A top panel). Subjects were fitted with an EEG cap, skin temperature and galvanic skin response sensors.

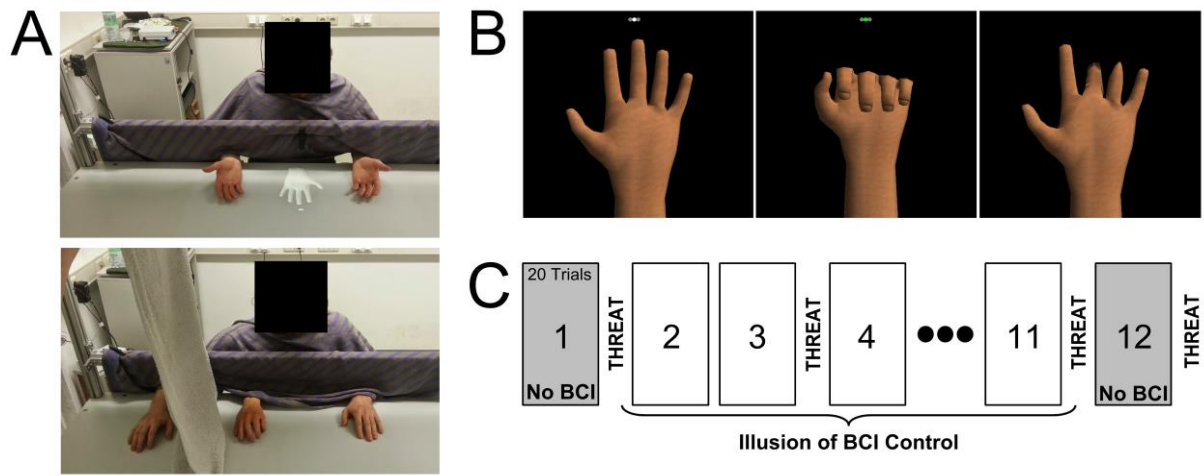

**Fig 1. Experimental methods.** A) The top panel shows the setup for our third hand illusion. The bottom panel shows our set up for the rubber hand illusion. B) The virtual third hand as seen by subjects during the third hand illusion. The leftmost and the middle panel show the starting and end position of the hand movement. The rightmost panel shows the 'broken' fingers at the threat. C) Block diagram of the experiment

Subjects were shown three cues corresponding to each hand, the cue colours indicated inter-trial interval (5s), preparation (2s) and go (3s) or no-go (3s). After the go cue subjects were instructed to close their hand (clench a fist) once and return to the rest position of an open hand palm up. For the real left and right hand this was a simple go/no-go cued movement execution in the ratio (80%:20% respectively). No-go trials were used to maintain the attention of the subjects and to keep movement vs no movement trials of the real hands in the same proportion as movement vs no movement for the virtual reality third hand (also 80%:20% respectively). When cued to control the virtual third hand subjects were instructed to perform a visual imagery of the hand closing until the trial was complete, "imagine seeing the hand close". Visual imagery was chosen as it may be less likely to interfere with the movement execution component of the task, whereas perhaps a kinaesthetic imagery would. It was explained to the subjects that the EEG cap they were wearing would allow the computer to recognise and classify features of the brain signals generated specifically by the imagery, and that if classified correctly this would generate a cued movement of the virtual reality hand. The hand played a single sequence to show it closing and opening a fist from a resting open position (Fig 1B left and middle panels). In reality however the EEG was not connected and positive feedback from the 'BCI controlled' trials (i.e. movement of the virtual hand) occurred randomly with a probability of 80%.

Subjects performed 12 blocks of 20 trials per block (Fig 1C), and were offered a short break in between each block. In the first and last block subjects were told there was no control of the hand via BCI and saw only trials of cued movement of the virtual hand. Subjects were asked to always perform the imagery they should generate for the experiment as described above (at the cue and continue while the virtual hand was moving) despite knowing the BCI was not functioning at this time. In the middle 10 blocks subjects performed 200 trials containing: 40 go trials and 10 no-go each for the left and right hands and 100 trials of cued movements of the virtual middle hand under 'BCI control' with 80% positive feedback. All trial types were randomly distributed across all 200 trials.

At the end of the $1^{st}$ (pre), $3^{rd}$ (early), $11^{th}$ (late) and $12^{th}$ (post) block a threatening stimulus was shown to the subjects to probe the illusion strength measured by the galvanic skin response (GSR). The 'early' response occurred at the third rather than the second block as we wanted that the result was not due to subject's naivety, but reflected the early response after some exposure to the task. The threatening stimulus consisted of a cracking sound while the virtual middle hand changed to show broken fingers (Fig 1B right panel).

## Rubber Hand Illusion - RHI

After the IBCI subjects then performed the rubber hand illusion [1]. Subjects remained seated in the same position as in the IBCI. In the rubber hand illusion experiment the right hand was shielded from view and a gender matched rubber right hand was placed on the table in front of the subject (Fig 1B bottom panel). The experimenter synchronously brushed the unseen right hand and the rubber hand following a brushing protocol consistent between subjects. Subjects were asked to comment throughout the experiment on the feeling of the illusion; firstly when they felt the illusion begin and then again if the illusion got stronger. Brushing continued for approximately 10mins apart from one case where the subject reported losing the illusion and brushing was stopped after 4mins.

## Rubber Third Hand Illusion - RTHI

After the RHI we performed the rubber third hand illusion (RTHI), described by Guterstam and colleagues [7]. The rubber third hand experiment was performed exactly as the RHI (above), with the single difference that subjects were able to see both their own hands in addition to the rubber hand, i.e. the real hand was not covered from view.

## Baseline Conditions

A subset of 8 subjects who performed all three illusion conditions returned for a separate session to perform baseline conditions of each illusion. To establish a baseline for the IBCI condition subjects first repeated only the first and last blocks of the IBCI as described above. Subjects made no movements, it was a static condition, but not passive as the movement imagery was still performed. Following this to establish a baseline for the RHI condition subjects repeated the RHI as described above, however in the first baseline case subjects first received 5 minutes of asynchronous brushing after which they answered the questionnaire. They then had a second baseline case with 5 minutes of synchronous brushing with the rubber hand rotated 90 degrees, viewed perpendicular to the subject's body after which they completed an additional questionnaire. Finally to establish a baseline for the RTHI condition subjects repeated the RTHI condition as described above, also with the two baseline conditions asynchronous and synchronous-rotated as described above.

## Data Acquisition and Analysis

Data acquisition and analysis was treated identically for both illusion and baseline conditions. The temperature was recorded during all conditions while the GSR only during the IBCI. Measurements were made always from both hands and all recordings were made using the actiChamp (Brainproducts GmbH, Germany) at 1kHz sampling rate. Skin temperature was recorded to address the aspect of disownership in the limbs during the illusion using a physiological measure [12]. Galvanic skin response was measured using the Brainproducts GSR Module to provide a physiological measure of ownership during the illusion conditions [7,13]. The two passive electrodes were taped with skin tape to the medial phalanx of the index and middle finger, which had been covered in conductive gel for GSR. The GSR response was calculated for each subject as the difference between the maximum conductance in a 5s window after a threat and the mean conductance over 5s before the threat. Skin temperature was measured using a Greisinger GMH3210 Digital thermometer adapted for the actiChamp. The thermocouple probe was taped to the skin 2cm behind the knuckle of the middle finger. The skin temperature was calculated as the difference between the hand that was matched in the illusion condition and the other hand [12]. A baseline correction was made by subtracting the mean difference in skin temperature for a 10s window around the subjects report that they first felt the illusion. For RHI and RTHI this report was given verbally; in the IBCI condition we used the time of the early threat, at this point we had the first evidence that subjects felt the illusion. The temperature data was resampled at a frequency of 1Hz and smoothed using a first order Savitzky-Golay filter, with a window width of 20s. The temperature and galvanic skin response data for one subject was not recorded due to a battery fault in the equipment; however the questionnaire responses for this subject were retained.

Subjects had to answer several questions immediately after each experiment. The questions were similar to those typically found in ownership/control illusion literature, but also were specifically designed to explore key differences between the feelings of ownership, control and independence in the three conditions we investigated. Subjects were asked to rate their answers using the seven-step visual-analogue scale ranging from strongly agree (+++) to strongly disagree (---).

Throughout our analysis we first assessed statistical significance across groups or conditions using a one-way non-parametric ANOVA (Kruskal-Wallis test). We then performed post-hoc pairwise Wilcoxon tests to assess statistical significance between pairs of groups or conditions. All reported p-values were corrected for multiple testing where applicable using the Holm-Bonferroni method and the adjusted p-values were reported. It is important to note that this method can produce adjusted p-values larger than 1.

# Results

Subjects performed three experiments (Fig 1): The imitation brain computer interface (IBCI) experiment was designed to induce the illusion of a BCI controlled third hand. In this experiment healthy human subjects were fitted with an EEG cap and told that by performing visual imagery in response to a cue they could move a virtual reality hand projected on a table top in front of them. Subjects could always see both of their own hands and cues to move the virtual hand were interspersed with cues to replicate the movement with their own hands. In reality the BCI was not connected and the virtual hand moved randomly with 80% probability. We compared our illusion to the classic Rubber Hand Illusion (RHI) [1] and the Third Hand Illusion (RTHI) [7]. All conditions were additionally compared to their baseline (see materials and methods).

## The illusion of ownership in IBCI

We first sought to determine whether our IBCI experiment gave subjects the feeling of ownership of a supernumerary hand. This was examined by means of questioning immediately after the experiment (Figs 2 and 3, Tables 1 and 2). Question (Q) 1 and 2 directly probe the feeling of ownership. For the IBCI condition significantly positive responses were recorded for Q1-2 compared to baseline (median (++), see Table 2 for p-values). In contrast for the RHI, subjects were disagreeing (median (---) for Q1 and Q2) not different from the baseline condition (Table 2). Also, subject's answers for RHI were significantly different from the IBCI condition (see Table 1 for p-values). This suggests a strong feeling of ownership of a supernumerary limb in the IBCI condition but little or no supernumerary ownership in the RHI and the baseline conditions. Between the IBCI and the RTHI we found a significant difference for Q2 with a stronger positive response for IBCI than for RTHI ((++) for IBCI, (+) for RTHI, Table 1) and no significant difference for Q1 (Table 1). Consistently, RTHI answers for Q1 and 2 were significantly more positive than during its baseline condition (Table 2). These results suggest that both our IBCI and the RTHI induce feelings of ownership of a supernumerary limb unlike the RHI, with the ownership feeling being stronger in IBCI than in RTHI.

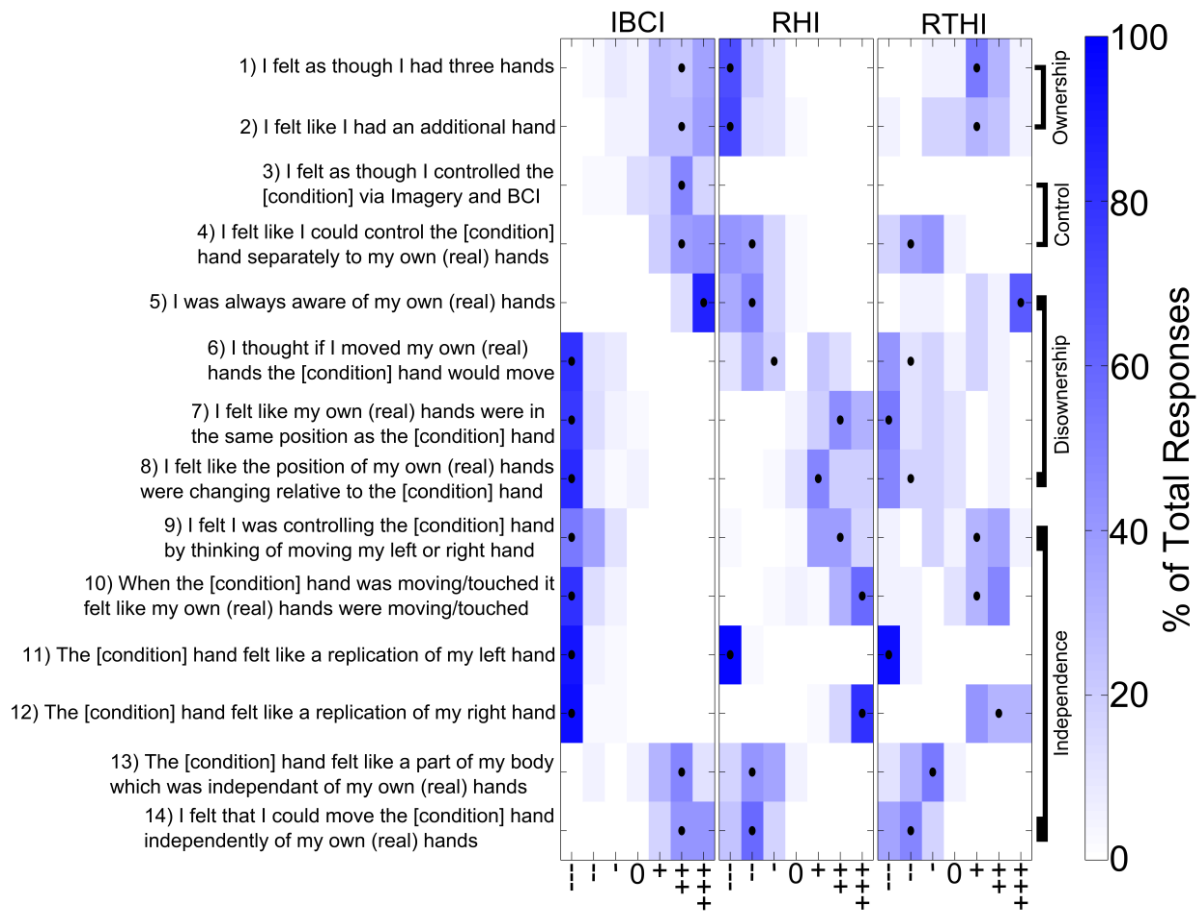

**Fig 2. The questions and responses for the different experimental conditions.**

Subjects were asked to complete the questionnaire immediately after each experimental condition using a seven-step visual-analogue scale ranging from strongly agree (+++) to strongly disagree (---). The black dots indicate the median responses and the colour scale indicates the percentage of the total responses. For the real questionnaires given to subjects the [condition] of each statement was substituted with "virtual" for IBCI and "rubber" for RHI or RTHI. See also Table 1 for statistical comparisons between different experimental conditions.

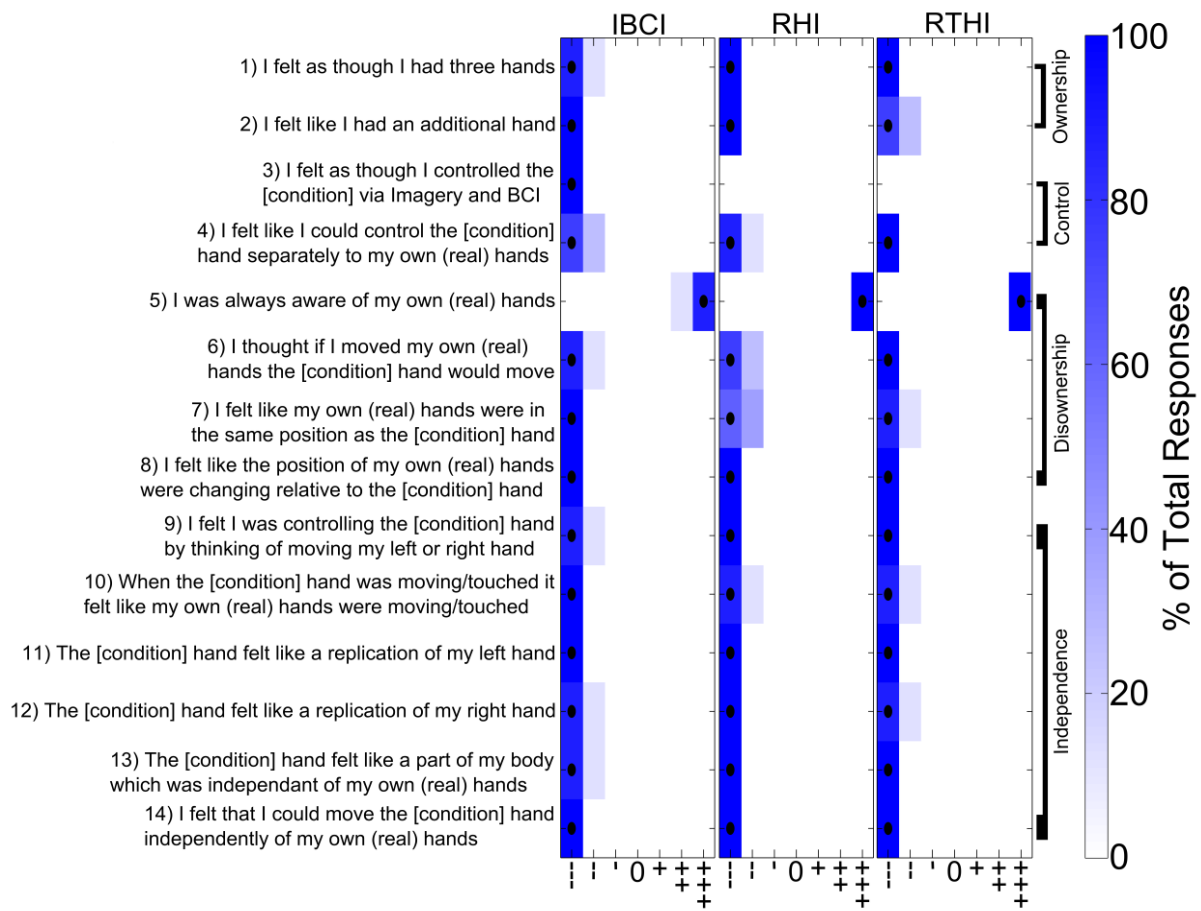

**Fig 3. The questions and responses for the different baseline conditions.** Subjects were asked to complete the questionnaire immediately after each baseline condition using a seven-step visual-analogue scale ranging from strongly agree (+++) to strongly disagree (---). The black dots indicate the median responses and the colour scale indicates the percentage of the total responses. For the real questionnaires given to subjects the [condition] of each statement was substituted with "virtual" for IBCI and "rubber" for RHI or RTHI. See also Table 2 for comparisons between baseline and illusion conditions.

**Table 1 - Statistical assessment of differences between conditions for each question.**

| Question No. and attribute addressed | Across all groups | IBCI vs RHI | IBCI vs RTHI | RHI vs RTHI |
|---|---|---|---|---|
| **1 (Ownership)** | $8.4 \times 10^{-15}$ | $4.7 \times 10^{-7}$ | 0.12 | $2.4 \times 10^{-9}$ |
| **2 (Ownership)** | $8.2 \times 10^{-15}$ | $3.8 \times 10^{-7}$ | $2.9 \times 10^{-3}$ | $7.2 \times 10^{-8}$ |
| **3 (Control)** | N/A | N/A | N/A | N/A |
| **4 (Control)** | $2.5 \times 10^{-15}$ | $2.8 \times 10^{-7}$ | $7.4 \times 10^{-9}$ | $3.0 \times 10^{-2}$ |
| **5 (Disownership)** | $5.4 \times 10^{-16}$ | $2.0 \times 10^{-7}$ | $3.7 \times 10^{-2}$ | $5.5 \times 10^{-8}$ |
| **6 (Disownership)** | $4.2 \times 10^{-8}$ | $4.3 \times 10^{-6}$ | $2.4 \times 10^{-3}$ | 0.14 |
| **7 (Disownership)** | $1.8 \times 10^{-15}$ | $2.3 \times 10^{-7}$ | $3.5 \times 10^{-2}$ | $6.4 \times 10^{-8}$ |
| **8 (Disownership)** | $3.4 \times 10^{-15}$ | $3.8 \times 10^{-7}$ | $6.0 \times 10^{-3}$ | $3.2 \times 10^{-7}$ |
| **9 (Independence)** | $1.2 \times 10^{-13}$ | $3.8 \times 10^{-7}$ | $2.0 \times 10^{-7}$ | $8.9 \times 10^{-2}$ |
| **10 (Independence)** | $3.7 \times 10^{-16}$ | $2.2 \times 10^{-7}$ | $1.0 \times 10^{-8}$ | $2.7 \times 10^{-5}$ |
| **11 (Independence)** | 0.59 | 1.5 | 1.2 | 1.5 |
| **12 (Independence)** | $2.7 \times 10^{-17}$ | $5.5 \times 10^{-8}$ | $3.1 \times 10^{-11}$ | $9.3 \times 10^{-5}$ |
| **13 (Independence)** | $3.4 \times 10^{-7}$ | $7.8 \times 10^{-4}$ | $1.2 \times 10^{-5}$ | 0.38 |
| **14 (Independence)** | $1.9 \times 10^{-8}$ | $5.2 \times 10^{-4}$ | $1.3 \times 10^{-6}$ | 0.61 |

Table 1 - P-values are given to 2 significant figures. Comparison across all groups was made using the Kruskal-Wallis test (column 2). Comparison of IBCI vs RHI uses Wilcoxon signed rank test, comparison between IBCI vs RTHI and RHI vs RTHI uses Wilcoxon rank sum test because not all subjects performed the RTHI. All p-values of the pairwise tests (columns 3-5) were corrected for multiple comparisons using the Holm-Bonferroni correction.

**Table 2 – Statistical comparison with baseline conditions.**

| Question No. and attribute addressed | IBCI vs IBCI baseline | RHI vs RHI baseline | RTHI vs RTHI baseline |
|---|---|---|---|
| **1 (Ownership)** | $2.3 \times 10^{-2}$ | 0.13 | $2.3 \times 10^{-2}$ |
| **2 (Ownership)** | $2.3 \times 10^{-2}$ | $6.3 \times 10^{-2}$ | $3.1 \times 10^{-2}$ |
| **3 (Control)** | $7.8 \times 10^{-3}$ | N/A | N/A |
| **4 (Control)** | $2.3 \times 10^{-2}$ | 0.13 | 0.13 |
| **5 (Disownership)** | 1.0 | $2.3 \times 10^{-2}$ | 1.0 |
| **6 (Disownership)** | 1.0 | $4.7 \times 10^{-2}$ | 0.5 |
| **7 (Disownership)** | 0.50 | $2.3 \times 10^{-2}$ | 0.5 |
| **8 (Disownership)** | 0.50 | $2.3 \times 10^{-2}$ | 0.5 |
| **9 (Independence)** | 0.38 | $3.1 \times 10^{-2}$ | $2.3 \times 10^{-2}$ |
| **10 (Independence)** | 1.0 | $2.3 \times 10^{-2}$ | $2.3 \times 10^{-2}$ |
| **11 (Independence)** | 3.0 | 3.0 | 2.0 |
| **12 (Independence)** | 1.0 | $2.3 \times 10^{-2}$ | $2.3 \times 10^{-2}$ |
| **13 (Independence)** | $2.3 \times 10^{-2}$ | $3.1 \times 10^{-2}$ | $3.1 \times 10^{-2}$ |
| **14 (Independence)** | $2.3 \times 10^{-2}$ | 0.13 | 0.13 |

Table 2 – A Wilcoxon signed rank test was used to compare the answers given in the experimental condition to the baseline condition for each illusion. All p-values were corrected for multiple comparisons using the Holm-Bonferroni correction.

To provide a further physiological measure of ownership during the IBCI condition the galvanic skin response (GSR) was recorded. An increased GSR response to a threatening stimulus is an established objective measure of ownership [7,13]. Threats were made to the virtual reality third hand at four time points during the experiment (Fig 4A). Either while the subjects believed they controlled the hand (early/late) or while subjects knew there was no control (pre/post) (see Materials and Methods). A significant difference in GSR was observed across all four threats (left hand $p=2.8 \times 10^{-10}$, right hand $p=7.2 \times 10^{-10}$, Kruskal-Wallis test corrected for multiple comparisons with Holm-Bonferroni correction). There was a strong and significant difference in the GSR response to a threat during the imitation BCI control (early/late) compared to the baseline conditions where subjects saw the virtual hand move but were told they are not controlling it with the BCI (pre/post) (left hand $p=3.4 \times 10^{-12}$, right hand $p=1.8 \times 10^{-12}$, Wilcoxon signed rank test, comparing between the combined pre/post and early/late threats, $p=0.36 \times 10^{-11}$ after correction for multiple comparisons with Holm-Bonferroni). While there were no significant differences for either hand between the pre and post conditions there was a significant increase in the GSR response from 'early' to 'late' threats (left hand $p=0.02$, right hand $p=0.004$, Wilcoxon signed rank test, correction for multiple comparisons with Holm-Bonferroni). This indicates that the ownership illusion became stronger over the course of the imitation BCI control. Furthermore, there is a comparably small influence of hand on the GSR: when the data for left and right hands was compared separately at each threat there was a weakly significant difference between the hands only at the 'late' threat ($p=0.04$, Wilcoxon sign rank test Holm Bonferroni corrected for multiple comparisons). Taken together, these results show that the strength of the ownership, as defined by the GSR, increases over time and this increase is restricted to only the period of imitation BCI control.

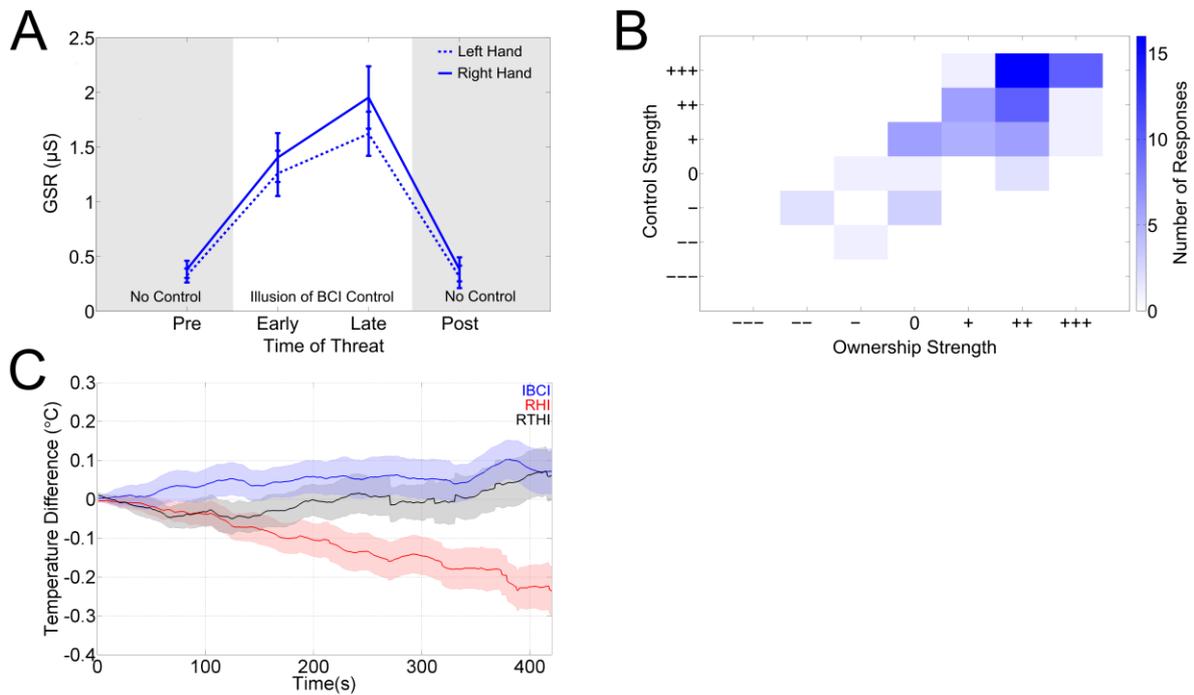

**Fig 4.** A) The mean across subjects GSR after a threatening stimulus during the IBCI condition. A significantly stronger response was seen during the imitation of BCI control compared to the pre and post conditions where subjects knew there was no control. Error bars indicate the standard error of the mean (SEM). B) Correlation between subject's responses to Q3 (addressing feeling of control) and Q1 and Q2 (both addressing the feeling of ownership) during the IBCI condition. As subjects answers for Q1 and Q2 were highly correlated within the IBCI condition (R=0.87, p<0.001, Spearman's rank correlation) the data from both questions has been pooled. Hence, for each subject the plot contains one paired sample representing the subject's answers to Q3 and Q1 and a second paired sample representing the answers to Q3 and Q2. The colour scale indicates the number of responses. C) Mean across subjects difference in temperature between the subjects real hands aligned to illusion onset (see Materials and Methods for details). Shaded areas show the SEMs.

## The illusion of control in IBCI

We assessed subjects feeling of control by questions Q3 & 4. Following the IBCI condition subjects answered all questions significantly positively compared to baseline conditions (median (++), Table 2), indicating a strong feeling of control. For RHI and RTHI the answers to Q4 were significantly different from the IBCI condition (Fig 2, Table 1), with subjects disagreeing to having control (median (--)), not different from baseline conditions (Table 2). This shows that there was little to no feeling of control during RHI and RTHI in contrast to IBCI. Q3 did not apply to the RHI and RTHI. In the baseline experiments subjects did not feel a sense of control in any of the conditions (Table 2, Fig 3). We recognise that the baseline experiments did not involve any overt movement. In the IBCI case subjects were performing movement imagery but were aware there is no control; we still used Q3&4 in this case to demonstrate that performing the imagery, viewing the hand and seeing it move in response to the cue did not contribute to the sense of control shown for the IBCI condition.

## The correlation between the illusion of control and ownership

Q3 was asked to determine how much subjects felt like they controlled the virtual reality third hand via BCI. While the median response was significantly positive compared to baseline ((++), Table 2), the answers of individual subjects spanned both positive and negative responses. Overall most answers were positive ($75^{th}$ percentile response +++, $25^{th}$ percentile response ++) but 7 out of 36 subjects gave answers which were negative or equal to 0. This indicates that some subjects may have realised that the BCI control was unrelated to their cognitive processes during the imitation 'BCI controlled' trials. Interestingly the strength of the illusion of ownership also varied across subjects with many subjects reporting a strong feeling of ownership while some did not report any feeling of ownership (Q1 & 2,

Fig 2). We therefore hypothesized that subjects with a strong illusion of control may also have a strong feeling of ownership and subjects without the illusion of control may not report a strong feeling of ownership. To investigate this we correlated the answers to Q3 (feeling of control) with the answers to Q1 and Q2 (feeling of ownership) and found strong and significant correlations (Fig 4B, R=0.64 between Q3 and Q1, R=0.73 between Q3 and Q2, $p<0.001$, Spearman's rank correlation). Importantly in no case did the subjects have the feeling of owning a third or additional hand without reporting some feeling of control via the BCI, or vice versa (Fig 4B). This shows that in the IBCI condition the illusion of ownership was closely linked to the illusion of control.

## Disownership of the real hands

To further explore the idea that the third hand was considered a supernumerary hand rather than the replacement of one of the real hands we investigated whether there was any disownership of the subject's real hands during the IBCI condition. Disownership refers to the loss of the awareness of ownership of one's own body parts, therefore no or little disownership of the subject's real hands would suggest that rather than replacing a limb, the illusion induces the feeling of a 'third hand'. We recorded skin temperature (see Materials and Methods) which has recently been shown to be a correlate of disownership of the real hand in the RHI [12]. We replicated the previously reported effect, as we observed a significant cooling of the matched hand between the mean of the $1^{st}$ and $6^{th}$ minute after the report of illusion onset in the RHI condition relative to the other hand (Fig 4C, $p=0.02$, Wilcoxon sign rank test, Holm-Bonferroni correction for multiple comparisons). We did not see this significant cooling in either our IBCI or the RTHI ($p=0.14$ and $p=0.97$ respectively, Wilcoxon sign rank test, Holm-Bonferroni correction for multiple comparisons). Comparing temperatures for the $1^{st}$ minute after the report of illusion onset there was no significant difference between the three conditions (for all conditions p=0.62, Kruskal-Wallis test. p=1.1 for IBCI vs RTHI and p=1.3 RHI vs RTHI, Wilcoxon rank sum test and p=1.3 IBCI vs RHI, Wilcoxon signed rank test, Holm-Bonferroni correction for multiple comparisons). When we compared the temperatures in those subjects who maintained the illusion into 6$^{th}$ min after the illusion onset there was a significant difference between the conditions (p=0.0019, Kruskal-Wallis test, the number of subjects was N(IBCI)=35, N(RTHI)=11 and N(RHI)=22); post-hoc tests showed significance between IBCI and RHI (p=0.002, Wilcoxon rank sum test, Holm-Bonferroni correction for multiple comparisons), boarder line significance between RTHI and RHI (p=0.056, Wilcoxon rank sum test, Holm-Bonferroni correction for multiple comparisons) and no significance between IBCI and RTHI (p=0.4, Wilcoxon rank sum test, Holm-Bonferroni correction for multiple comparisons). The average duration for the illusion was 7.86min +- 19.0s in RHI and 7.33mins +- 20.8s in RTHI (mean +- SEM, 7.6s over both). During the IBCI subjects were asked not to speak, so did not report the specific onset or loss of the illusion during the experiment, this was in order to maintain the conditions of a typical BCI experiment. These results are in support of previous findings [12], suggesting that there is little to no disownership of the subject's actual hands during the IBCI and RTHI condition while there are signatures of disownership during the RHI.

In addition disownership was investigated by questions Q5-8 (Fig 2). Q5 specifically addressed the awareness subjects had of their own hands. Whilst all responses were significantly different across experiments (Table 1) the conditions IBCI and RTHI showed strongly positive results (median (+++)), not different to baseline (Table 2) indicating no or less disownership than during RHI where a negative response was measured (median (--)), which was significantly lower than during baseline (Table 2). In the baseline conditions subjects gave strongly positive results for all illusions (median (+++), Fig 3) indicating that there was no feeling of disownership during the baseline conditions. We further explored subjects' disownership of their own hands by Q6-8. These questions were again answered significantly differently for each of the three conditions IBCI, RHI and RTHI (Table 1). For the IBCI all questions had strong negative responses (median (---)), which were not different from baseline (Fig 3, Table 2), indicating full awareness of the real hands during this condition. For the RTHI responses were also negative but with broader distributions (Fig 2), though neither different from baseline (Fig 3, Table 2). For the RHI Q7-8 showed positive responses but a negative response was recorded for Q6, in all cases the RHI response was significantly different to its baseline (Table 2). The findings from the questionnaire therefore corroborate that of the temperature data, indicating that there is no disownership of the real hands during the IBCI condition in contrast to clear disownership during the RHI. During the RTHI condition there seems to be no or little disownership, however the RTHI shows a significantly weaker and broader response range compared to the IBCI, suggesting a less clear absence of disownership than during the IBCI. Next we investigated if the supernumerary hand was also felt to be independent from the real hands during IBCI.

## The independence in IBCI

To assess independence, we used several questions (Q9-Q14) which specifically addressed the feelings of independence both in terms of ownership and control. Q13 references independent ownership and showed significantly more positive answers in the IBCI than during its baseline (Table 2). This suggests a feeling of independent ownership for the IBCI condition. RHI and RTHI answers had a negative median, which was only slightly but still significantly higher than during their baselines (Fig 3, Table 2). This suggests some feeling of independent ownership also for RHI and RTHI, however, less pronounced than for IBCI where the answers were significantly more positive than in RHI or RTHI (Table 1).

Questions 11-12 addressed the extent to which the artificial hand felt like a replication i.e. a less independent limb. For the IBCI condition (where a left virtual hand was projected) the answers to Q11, probing the replication of the subjects' own left hand, were negative (median (---)) and not significantly different from baseline (Table 2), whilst for the RHI and RTHI (where a right rubber hand was used) the answers to Q12, probing the replication of the subjects' own right hand, were significantly positive compared to baseline (median (+++) and (++) respectively, Table 2). This shows that the feeling in the RHI and RTHI was a replication of the existing hand, while the IBCI condition created the feeling of a separate hand.

The feeling of independent control was assessed with Q14 and was answered significantly more positively in the IBCI condition compared to the RHI or RTHI (Table 1), with no significant difference between RHI and RTHI (Table 1). The strongly negative responses to this question for the RHI an RTHI where not different to baseline (median (--), Table 2). In contrast the positive responses during IBCI (median (++)) were significantly different to baseline (Table 2) suggesting that the feeling of being able to control the hand independently to the real hands was only felt in the IBCI condition. Q9-10 additionally probed the independence of control with positive answers indicating less independence. The positive responses reported for the RHI and RTHI condition compared to baseline (median (++) and (+) respectively, Table 2) and the negative responses for IBCI, not different from baseline (median (---), Table 2) confirmed that the control was only felt as being independent in the IBCI condition (see also Table 1 demonstrating the significance of the difference between IBCI and RHI/RTHI). This conclusion is further corroborated by Q10 where all conditions were significantly different (Table 1), with strong support for independence in the IBCI condition (median (---), not significantly different from baseline, Table 2) and no support for independence in the RHI and RTHI conditions (median (+++) and (+) respectively, significantly different from baseline, Table 2).

## Baseline conditions

A subset of 8 subjects who performed all three illusion conditions were called back at a later date to perform the baseline conditions (see Materials and Methods for details). The baseline conditions were performed to establish a baseline measure of responses to assess the presence/strength of the illusions in the 'illusion conditions'. After each baseline experiment subjects were asked to answer the questionnaire again. These responses were compared to the responses given in the original experimental conditions (Table 2). Not a single subject reported an illusion in any of the three baseline conditions and the answers to the questions were very consistent across subjects (Fig 3). In the RHI and RTHI condition subjects were asked to answer questions separately for the asynchronous and the synchronous-rotated brushing conditions. Only 3 subjects answered cumulatively 5 questions differently in the rotated condition (with a maximum difference of one step on the visual analogue scale per questions in 4/5 cases, and one with a difference of 2 steps). The difference never changed the sentiment of the subjects answer, highlighting the equal ability of each baseline condition to remove any illusion of ownership and agency. For this reason the results presented in Fig 3 are for the asynchronous baseline questionnaire. Furthermore subjects were asked additional questions after the baseline conditions: 1) It felt as if my real hand were matching the texture of the [condition] hand. 2) I felt as if the illusion I was experiencing was coming from somewhere between my real hand and the [condition] hand. For these questions all subjects gave a '---' response.

# Discussion

We present the IBCI condition that induces the illusion of ownership in an independent supernumerary limb and the sense of agency for its actions. To distinguish this new illusion from existing body ownership illusions we compared it to the rubber hand illusion (RHI) [1]. and a variant of it, the rubber third hand illusion (RTHI) reported to give the feeling of ownership in a third hand [7].

## The IBCI illusion – Ownership, Control and Independence

The IBCI illusion combines the illusion of ownership of an independent third hand and agency of its movements, with no disownership in the real hands. Our illusion demonstrates in able-bodied subjects the brains capacity to expand the body representation beyond the body's anatomy by a supernumerary limb that is not just a replication of an existing body part [5–7,14] .

Questionnaire and GSR results demonstrated that subjects felt both a strong sense of ownership and control over the third hand, which was not affected by the sight or movement tasks performed with their real hands during the experiment. It has recently been suggested [12] that the skin temperature of a hand which is disembodied in the illusion cools relative to the other hand [15,16]. While this effect seems to be weak, it offers a physiological measure of disownership. Our results replicated the cooling effect in skin temperature during the RHI (Fig 4C). Importantly the cooling is not observed in our IBCI or the RTHI, which suggests no disembodiment in these conditions. We addressed the feeling of independence in the ownership and control and found a noticeable difference between the IBCI and the RTHI:

subject's responses indicated strong feelings of independence for the IBCI and little to no independence in the RTHI. Furthermore there was no significant difference in the independence between the RHI, where we demonstrate disembodiment, and the RTHI. Our results, therefore, indicate that the IBCI induces the illusion of an independent supernumerary limb while the RTHI seems to feel more like a replication of the existing hand. Previously the feeling of independent ownership has only been shown in amputees [8], where three out of fourteen subjects reported that their arm shown in a mirror box felt like an additional limb, independent to their feeling of the phantom limb.

We were able to induce the illusion using an imitation BCI, showing the correct movement on 80% of the trials, with no causal connection to subject's brain activity. This is fundamentally different to previous ownership illusions, including both the RHI and RTHI, which either rely on combined visual and tactile feedback or on actual BCI control, i.e. combined motor signals and visual feedback. Subjects in the IBCI illusion received only visual feedback in response to a cued task, unrelated to their brain activity.

## Relationship between control and ownership

During the IBCI condition most subjects believed they controlled the arm via the BCI, and in all these cases the illusion of control accompanied the illusion of ownership (Fig 4B). Importantly in no subjects was one felt without the other and across subjects the strength of the illusion of ownership was correlated with the strength of the illusion of control. The relationship or causality between the feelings of ownership and the sense of agency remain unclear, and some studies suggest they represent two distinct cognitive processes [17]. We have demonstrated here that the illusion of ownership and control can come from a task where the subject has no actual control. This supports the notion that success regardless of actual influence in a task [18] and observing expected congruent movements or feedback can lead to the feeling of control [19,20], but extends this notion to ownership and additionally to supernumerary limbs. In the IBCI condition, subjects received with 80% probability a positive feedback, i.e. in 80% of the trials the virtual hand moved during instructed movement imagery. It remains an open question exactly how accurate the performance must be to maintain the illusion? For example, if subjects were to receive 80% negative feedback and only 20% successful trials would they still feel the same sense of ownership and agency, or would subjects report no sense of ownership and agency as in the baseline condition where they know there is no control. The sense of agency has been shown to increase with the congruency of the expected outcome during a BCI task [21]. How the sense of ownership is affected remains to be addressed by future investigations.

In the IBCI condition the illusion of ownership as demonstrated by the GSR was present only during the perceived BCI control (Fig 4A). Even though subjects were asked to perform the exact same imagery in response to the cue in the pre and post blocks, the GSR was significantly lower compared to IBCI - the only difference being that here they were told there is no control. Hence, the illusion of ownership was independent of motor imagery but tied to the belief of control, in keeping with a recent study [21] reporting that the sense of agency in BCI is independent of motor imagery-based neural signals and dominated by visual feedback.

## Relevance for brain-computer interfaces

The motivation for exploring a "third hand" illusion arises from the likely concept that BCIs could also be used to supplement existing function [22], for example by providing control of supernumerary limbs which could be used concurrently to movements of the real limbs. Our study suggests that in such BCIs subjects may feel ownership of the BCI controlled limb without disownership of the real limbs. BCI learning and control has been associated with the formation of stable cortical maps [23], and in addition several studies [24–27] suggest that brain areas and features of brain signals typically used for BCI control can be modified as a result of altering body ownership. It remains an open and interesting question as to whether a BCI controlling a supernumerary limb could induce a robust independent neural representation of the external effector.

In summary we present a new illusion that induces the feeling of independent ownership and control of a supernumerary hand. The illusion is induced by imitating the behaviour of a brain-computer interface, where subjects believe they control the movements of a virtual hand by their brain activity while in reality there is no causal link between subjects' brain activity and the movements of the hand.